# Robust Centerline Prediction for Accurate Vessel Wall Visualization of Intracranial Vessels in Multi-Contrast 3D MRI Data

P. Vogel, T. Kampf, K. Guggenberger, E. Raithel, C. Forman,
S. Meckel, U. Ludwig, A.J. Krafft, J. Hennig, T.A. Bley

*Abstract*—**An accurate planar visualization (curved planar reformation – CPR) of intracranial arteries is mandatory for an improved luminal and mural assessment especially in low resolution Magnetic Resonance Imaging (MRI) data sets acquired in standard clinical settings. CPR visualization methods based on the centerline of the desired structure are fast and easy to implement but the accuracy strongly depends on the spatial resolution of the 3D data set and the size of the desired vessel. In the manuscript, a novel algorithm for fast and robust centerline calculation in multi-contrast 3D MRI data is presented. It considers the extracted surface of the vessel structure for a more accurate centerline prediction resulting in an enhanced CPR visualization of small vessels.**

*Index Terms*— **centerline extraction, multi-contrast, VWI, vessel wall imaging, 3D visualization, CPR.**

## I. INTRODUCTION

MODERN Magnetic Resonance scanning technologies, such as MRI (Magnetic Resonance Imaging) have facilitated non-invasive imaging of the intracranial vasculature. Clear and artifact free depiction of the intracranial vessels is paramount for assessment of luminal and mural vascular pathologies such as atherosclerosis and vasculitis [1]. In order to precisely measure mural thickness or luminal stenosis of an intracranial artery a true orthogonal view on the evaluated vessel segment is mandatory. This can be achieved by utilizing curved planar reformation (CPR). CPR is an important technique in medical imaging to pro-vide a planar visualization of tortuous tubular structures such as blood vessels [2, 3], e.g., for vessel wall imaging of intracranial arteries [4, 5].

Therefore, longitudinal cross-sections are generated to display the lumen, the vessel wall and the surrounding tissue in a curved plane. Several approaches for calculating CPRs have been demonstrated in the past, whereas all methods strongly depend on the accuracy of the extracted centerline (3D-path) through the tubular structure [6]. For the extraction of the centerline of a desired 3D structure, different approaches have been presented, such as skeletonizing of binary 3D data [7] or tracking-based techniques [8]. While the latter approach requires a complex mathematics and difficult software implementation, the first approach is more sensitive to noise in the data. However, the easy implementation of the skeletonizing/thinning algorithm and additional 3D Dijkstra algorithm for binary vessel extraction [9] as well as the easy-to-handle and robust data processing, makes this combination preferable for a flexible method.

As an example, an MRI data set (time-of-flight sequence) of the internal carotid artery (cavernous segment to terminus) is shown in Fig. 1. After applying 3D Dijkstra (Fig. 1 (b)) and thinning algorithms [7], a single voxel chain remains representing the centerline of the binary 3D data set (Fig. 1 (c)). The accuracy of the centerline depends on the resolution of the given data set, which can yield large deviations from the real center of the structure especially for small vessels structures in

This work was supported in part by a grant from Deutsche Forschungsgesellschaft (DFG) under grand numbers HE 1875/26-1, BL 1132/1-2, and VO 2288/1-1.

P. Vogel is with the Department of of Diagnostic and Interventional Radiology, University Hospital Würzburg, 97080 Würzburg, Germany and with the Department of Experimental Physics 5 (Biophysics), University of Würzburg, 97074 Würzburg, Germany (e-mail: Patrick.Vogel@physik.uni-wuerzburg.de).

T. Kampf is with the Department of of Diagnostic and Interventional Neuro-radiology, University Hospital Würzburg, 97080 Würzburg, Germany and with the Department of Experimental Physics 5 (Biophysics), University of Würzburg, 97074 Würzburg, Germany (e-mail: Thomas.Kampf@physik.uni-wuerzburg.de).

K. Guggenberger is with the Department of of Diagnostic and Interventional Neuroradiology, University Hospital Würzburg, 97080 Würzburg, Germany (e-mail: Guggenberger_K@ukw.de).

E. Raithel is with Siemens Healthcare GmbH, 91058 Erlangen, Germany (e-mail: Esther.Raithel@siemens-healthineers.com).

C. Forman is with Siemens Healthcare GmbH, 91058 Erlangen, Germany (e-mail: Christoph.Forman@siemens-healthineers.com).

S. Meckel is with the Department of Neuroradiology, Medical Center, University of Freiburg, Faculty of Medicine, University of Freiburg, Freiburg, Germany (e-mail: Stephan.Meckel@uniklinik-freiburg.de).

U. Ludwig is with the Department of Radiology, Medical Physics, Medical Center, University of Freiburg, Faculty of Medicine, University of Freiburg, Freiburg, Germany (e-mail: Ute.Ludwig@uniklinik-freiburg.de).

A.J. Krafft is with the Department of Radiology, Medical Physics, Medical Center, University of Freiburg, Faculty of Medicine, University of Freiburg, Freiburg, Germany and with Siemens Healthcare GmbH, 91058 Erlangen, Germany (e-mail: axeljoachim.krafft@siemens-healthineers.com).

Jürgen Hennig is with the Department of Radiology, Medical Physics, Medical Center, University of Freiburg, Faculty of Medicine, University of Freiburg, Freiburg, Germany (e-mail: Juergen.Hennig@uniklinik-freiburg.de).

T.A. Bley is with the Department of Diagnostic and Interventional Neuroradiology, University Hospital Würzburg, 97080 Würzburg, Germany (e-mail: Bley_T@ukw.de).

low-resolution 3D data. To overcome this issue, interpolation algorithms, e.g., Bézier interpolation [10], can be used to calculate a smooth 3D-path with sub-voxel resolution based on the calculated voxel chain (Fig. 1 (d)). Unfortunately, for small vessels diameter compared to the spatial resolution of the data set, the deviations from the real vessel center are still not negligible, which yields im-perfections in the curved planar reformation (CPR) view [3] as indicated in Fig. 1 (e).

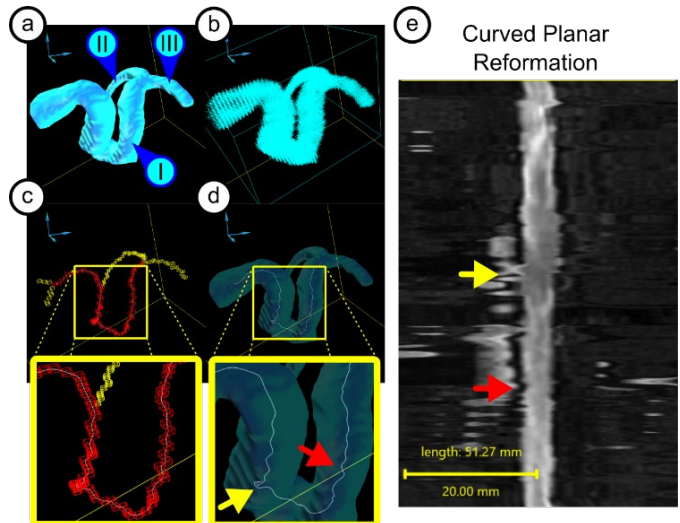

Fig. 1: Overview of the steps for a curved planar reformation (CPR) of the internal carotid artery (I). The lumen of a desired vessel structure, here indicated as 3D mesh model **(a)**, is voxel-wise extracted using the 3D Dijkstra algorithm **(b)**. After applying a thinning algorithm, a single voxel chain remains representing the central axes through the structure with an accuracy of the given data set **(c)**. To improve the accuracy of the real centerline, interpolation algorithms can be applied to calculate a 3D-path with sub-voxel resolution **(d)**. Based on this 3D-path, a straightened CPR view can be calculated **(e)**, which shows inaccuracies due to the voxel resolution. The red and yellow squares indicate different segments representing different parts of the structure (A. cerebri media (II), A. cerebri anterior (III)).

The objective of this work is to present a fast and robust approach to enhance the centerline prediction of an extracted volume from multi-contrast MRI data. For that, a correction step is added considering additional information of a calculated 3D mesh surface of the vessel structure, which can be determined by marching cube algorithm [11].

## I. MATERIAL AND METHODS

### A. Magnetic Resonance Imaging Sequences

For evaluation of the proposed centerline prediction approach, two MRI sequences have been used to generate the 3D data on a 3T whole-body scanner (MAGNETOM Prisma, Siemens Healthcare, Erlangen, Germany) with 64-channel head coil. Additional to a standard time-of-flight (TOF) sequence providing positive vessel contrast, a sagittal compressed-sensing (CS) accelerated T1-weighted SPACE prototype (Compressed Sensing - Sampling Perfection with Application optimized Contrasts using different flip angle Evolution) [12] has been implemented providing a dark-blood contrast.

The imaging parameters for the 3D TOF sequence are TE/TR=3.69/22 ms, averages=1, bandwidth=185 Hz/Px, flip angle=18°, slice thickness=0.6 mm, resolution=0.45×0.45×0.6 mm³, FOV=180×200×100 mm³, acquisition time=1:17 min, reconstruction matrix=404×448×168 px³.

The imaging parameters for the CS-SPACE sequence are TR/TE=800/5.1 ms, FOV=210×210×140 mm³, matrix=384×384×256 px³, slice thickness=0.6 mm, in-plane resolution=0.55×0.55 mm², pixel-bandwidth=450 Hz/px, flip angle=120°, ETL=50, acquisition time=6:47 min (k-space undersampling factor 0.22 resulting in 5-fold acceleration.

### B. Conventional Data Processing

In Fig. 2, the conventional workflow is sketched starting from a 3D MRI data set to curved planar reformation (CPR) view (black arrows). In addition, the steps for the proposed correction approach is shown in the grayed area following the blue arrows.

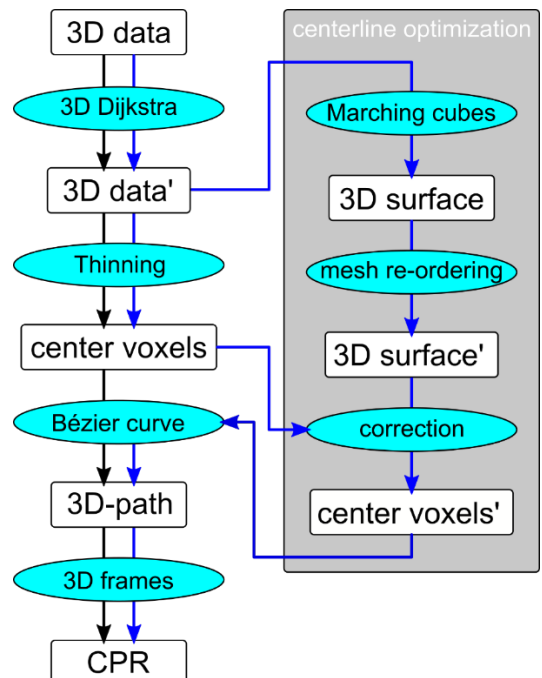

Fig. 2: Sketch of the data processing from 3D MRI data to CPR view (the black arrows indicate the standard workflow way and the blue arrows indicates the proposed correction). In a first step, the desired lumen is extracted using 3D Dijkstra algorithm resulting in a binary 3D data set (3D data'). **Left (black arrows)**: After applying thinning algorithm, the resulting central voxel line is used as basis for calculating a 3D-path (e.g., utilizing Bézier interpolation). Finally, multiple 3D frames are calculated along the 3D-path and used for CPR visualization. **Right (blue arrows)**: For optimization of the centerline prediction (3D-path), a meshed surface is calculated from the binary 3D data using Marching cube algorithm. After re-ordering of the triangle data, the optimized surface (3D surface') is used for calculating optimized center voxels (central voxels') to smooth the 3D-path through the vessel structure.

In a first step, the MRI data set is converted from the commonly used DICOM (Digital Imaging and Communications in Medicine) format [13] to the open-source format NIfTI (Neuroimaging Informatics Technology

Initiative) [14] and is stored in a normalized [0..1] three-dimensional array with dimension [*dimX*, *dimY*, *dimZ*] of floating-point numbers. Using a home-built software [12, 15] providing 3D visualization and a user-friendly graphical user inter-face (GUI), the desired vessel structure can be selected from the data by defining seed-points. From these points the inner lumen of the vessel structure is extracted using 3D Dijkstra [9] algorithm based on multiple intensity thresholds [12, 15] resulting in a binary 3D data set (3D data') [0, 1] with the size [*dimX'*, *dimY'*, *dimZ'*]. Applying thinning algorithm [7], a central single voxel chain (centerline voxels) with end-points, through-points and branch-points can be calculated representing the lines through the center of the vessel structure [15, 16]. Between an end-point and a branch-point or another end-point, single segments consisting of $N_{voxels}$ voxels $s_i$ are calculated representing different parts of the structure. The software can visualize one or multiple connected segments (cf. Fig. 1).

Using these center voxels $\{s_i\}$, the desired segment consists of, a smooth 3D-path (centerline) can be calculated utilizing, e.g., Bézier interpolation algorithm [10]. The calculated smooth 3D-path (centerline) consists of a user-defined number of points $N_{3D\text{-path}}$ serving as starting points for 3D frame calculation ($N_{frame}=N_{3D\text{-path}} \geq N_{voxels}$) required for CPR view preparation [4, 17].

One drawback of this approach, especially for low-resolution data, is the inaccuracy of the 3D-path due to the fact, that the underlying voxel chain can show large leaps between adjacent voxels in contrast to the dimensioning of the vessel structure, depending on the spatial resolution of the given data set (see Fig. 3). As a result, the 3D-path can show a step shape corrupting the CPR view.

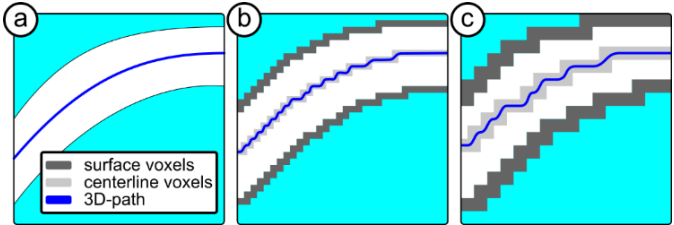

Fig. 3: Centerline prediction in data sets with different resolutions: **(a)** very high resolution, **(b)** medium resolution and **(c)** and low-resolution. Based on the centerline voxels, the 3D-path is calculated. With decreased resolution, the step shape of the centerline becomes clearly visible.

### C. Centerline Optimization

For a better prediction of the centerline (3D-path), additional steps are proposed. First, the surface information (Fig. 3 – surface voxels) of the extracted vessel structure is considered. The surface voxels represent the outer voxel shell around the extracted 3D vessel structure (binary 3D data'). To calculate a surface mesh covering the entire vessel structure, a marching cube algorithm [12] is applied generating a mesh consisting of multiple triangles $t_i$, each spanned by three independent points (vertices) $t_j = \{r_{3 \cdot j}, r_{3 \cdot j+1}, r_{3 \cdot j+2}\}$. Thus, the entire set of surface vertices $v_{MC} = \{r_i\}$ consists of $3 \cdot N_{triangle}$ points (three points per triangle), where $N_{triangle}$ means the number of triangles the surface mesh consists of.

For each segment point $s_i$ of the centerline predicted by the standard workflow, a sphere with a user-defined constant radius $R_{sphere}$ is created, which should be larger than the radius of the vessel structure $R_{sphere} > R_{vessel}$.

In a second step, the spatial information of the surface triangle points (vertices) $r_j$, which are located inside the sphere, are used for interpolating a new segment point $s'_i$ creating an optimized set of optimized center voxels $\{s'_i\}$ with sub-voxel resolution

$$s'_i = \frac{1}{N_{voxels}+1} \cdot \left( s_i + \sum_{j=1}^{N_{voxels}} r_j \right), \text{with } N_{voxels}$$
$$= |\{s_i | R_{sphere} > \|r_j - s_i\|\}|. \quad (1)$$

Eqn. (1) can be applied in two different ways with and without the initial center voxel $s_i$. In Fig. 4 (a), the sketch shows the virtual sphere around a segment point $s_i$ with radius $R_{sphere}$. The calculated frame with vectors $\{u, v, t\}$ spans up a local coordinate system oriented along the tubular structure [17]. The projected slice spanned up by $\{u, v\}$ (Fig. 4 (b)) indicates the new segment point $s'_i$ (center voxels*) calculated with the points $r_j$ from the including vessel surface vertices (green dots).

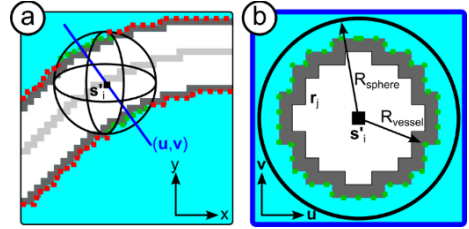

Fig. 4: Sketch of the optimization step: **(a)** each segment point $s_i$ is used as center of a sphere selecting specific vertices $r_j$ on the vessel surface to calculate a new center voxel $s'_i$ for an optimized 3D-path. The green and red dots indicate included and excluded vertices. **(b)** shows the cross-section through the tubular structure.

The proposed step can require high amount of computation time, especially for large number of vertices. Furthermore, a rough surface structure or protuberances can cause an erroneous centerline prediction. To reduce the number of vertices of the mesh surface and smooth the surface of 3D mesh, a refinement of the mesh (triangles) have to be performed. For that, an interpolation algorithm has been applied to calculate an interpolation between adjacent triangles $t_{adj}$ on the surface for each vertex $r_i$. The new vertex $r'_i$ is the mean of all vertices that belong to the adjacent triangles of $r_i$ without $r_i$ itself taking their multiplicity into account:

$$r'_i = \frac{1}{2 \cdot N_{t_{adj}}} \cdot \sum_{t_{adj}} \sum_k r_k \text{ with } \{r_k \in adj \text{ triangles} \wedge r_k$$
$$\neq r_i\}. \quad (2)$$

The goal is to find the adjacent triangles in a fast way. Since the 3D mesh data after Marching cube algorithm are stored in an unsorted array, it is useful to re-order the entire data set. This step allows to overcome both issues and decreases the computation time for collecting the points $r_i$ and also improves the surface structure (interpolation).

### D. Mesh Re-ordering

The computational effort of the proposed optimization step can be high due to the large number of surface vertices. In addition, the array of surface vertices $v_{MC}$ calculated by the marching cube algorithm consists of three times more vertices as necessary (for a closed surface mesh) since adjacent triangles $t_i$ are represented by different vertices $r_j$ with the same coordinates. To overcome this issue, the data of the surface mesh have to be re-ordered and optimized using two separate lists. The first list, the vertexlist $vl=\{r_i\}$, consists of all vertices $r_i$ of the mesh, without duplicated points, whereby the instruction for selection follows algorithm 1. The parameter $\varepsilon$ serves as interpolation parameter and allows combining all vertices within a specific radius (epsilon). This results in a less rough surface and reduces the number of surface triangles as indicated in Fig. 5.

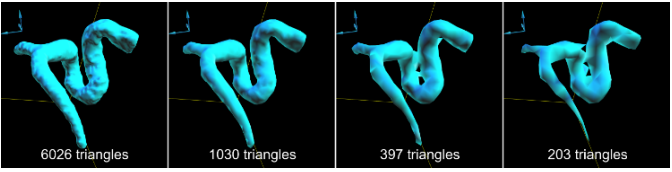

Fig. 5: 3D surface mesh with different epsilon factor ($\varepsilon_1$=1, $\varepsilon_2$=2, $\varepsilon_3$=3, $\varepsilon_4$=4) yields a less rough surface but also with reduced amount of triangles (surface interpolation).

```
Algorithm 1 point-selection is

input:  array of vertices v_CM={r}
        epsilon ε
output: array of vertices vl={r'}

vl := {}

for each entry r_i in v_CM do
    for each entry r_i' in vl do
        if ‖r_i' - r_i‖ > ε then
            add r_i to vl

    return vl
```

The second list, the indexlist $il=\{j\}=\{(i_1, i_2, i_3) \cdot N_{triangles}\}$, consists of all triangle information, which means for each triangle $t_j$ three indices $i_1=3 \cdot j$, $i_2=3 \cdot j+1$ and $i_3=3 \cdot j+2$.
For the given square example shown in Fig. 6 built with $N_{triangle}$=8 triangles, the array of vertices $v_{CM}$ is given by

$$v_{CM} = \{p1, p2, p5, p1, p5, p4, p2, p3, p6, p2, p6, p5,\\ p5, p6, p9, p5, p9, p8, p4, p5, p8, p4, p8, p7\}. \quad (3)$$

Multiple vertices appear several times within the list, which on the one side requires more memory capacity (here: $|v_{CM}|=N_{triangle}\times 3\times 3\times$double=576 Bytes), and on the other side more computational effort for further processing, such as surface interpolation.
The representation utilizing a vertexlist $vl$ and an indexlist $il$ for the square example is given by

$$vl = \{p1, p2, p3, p4, p5, p6, p7, p8, p9\} \quad (4)$$
$$il = \{1, 2, 5, 1, 5, 4, 2, 3, 6, 2, 6, 5, 5, 6, 9, 5, 9, 8, 4, 5, 8, 4, 8, 7\},$$

with a memory size of $|vl|$=9×3×double=108 Bytes and $|il|=N_{triangles}\times 3\times 1\times$integer=96 Bytes. The vertexlist $vl$ consists of all vertices of the 3D model and the indexlist $il$ defines the triangles by indexing the corresponding vertices in $vl$.
For a higher flexibility, the vertexlist $vl$ stores additional information about the indexed triangles $t_i$ in a double-linked list. For each entry $i$ of the vertexlist $vl$, an additional sub-list $il_{sub}$ is created, which stores all indices $j$ (entries) of that vertex $i$ within the indexlist $il$.
For the given example of the square, each entry {1..9} of the vertexlist $vl$ stores the information about the vertex position (**vert**=) as well as the information about the corresponding indices in the indexlist il ($il_{sub}$=) indicating the use of this vertex in the 3D model

$$vl(1) = \{vert = p1, il_{sub} = \{1, 4\}\}$$
$$vl(2) = \{vert = p2, il_{sub} = \{2, 7, 10\}\}$$
$$vl(3) = \{vert = p3, il_{sub} = \{8\}\}$$
$$vl(4) = \{vert = p4, il_{sub} = \{6, 19, 22\}\}$$
$$vl(4) = \{vert = p4, il_{sub} = \{6, 19, 22\}\}$$
$$vl(6) = \{vert = p6, il_{sub} = \{9, 11, 14\}\}$$
$$vl(7) = \{vert = p7, il_{sub} = \{24\}\}$$
$$vl(8) = \{vert = p8, il_{sub} = \{18, 21, 23\}\}$$
$$vl(9) = \{vert = p9, il_{sub} = \{15, 17\}\}, \quad (5)$$

with additional memory requirement of 24×1×integer=96 Bytes.

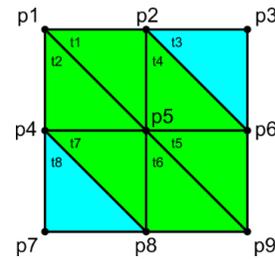

Fig. 6: Mesh square example consisting of 8 triangles {$t_1..t_8$} and 9 vertices {$p_1..p_9$}. After re-ordering of the mesh data, an interpolation of a vertex (here $p_5$) can be rapidly calculated (green area) by finding neighboring vertices.

### E. Surface Mesh Interpolation

After re-ordering of the mesh data, a surface interpolation can be calculated rapidly using the double-linked lists to find adjacent vertices. For that, the additional index-array $il_{sub}$ of the desired vertex $r_i$ is used to find adjacent triangles and vertices. Algorithm 2 demonstrates the entire process: for each index-entry $j$ within the entrylist $il_{sub}$ in the vertexlist $vl(i)$ of the desired vertex $i$, the position within the triangle ordering triple

(1, 2, 3) is calculated to find the right index for picking the vertex within the vertexlist $vl(j+di)$. Calculating $j$ mod 3 provides the positioning within the triangle ordering, e.g., for $j$ mod 3=1, the first vertex of the triangle triple is meant and thus for interpolation the other two ($j+1$ and $j+2$) have to be used.

For the given square example and a desired index $i=5$, the interpolated vertex $p'$ is calculated by summing up adjacent vertices $p1$, $p2$, $p6$, $p9$, $p8$, and $p4$ (example for the first entry $j=3$ → $j$ mod 3=0 → $d1=vl(j-2)=p1$ and $d2=vl(j-1)=p2$).

As a result, the surface of the mesh can be smoothed and undesired steps and unevenness can be interpolated, which results in an enhanced visualization of the vessel surface and an optimized centerline prediction.

```
------------------------------------------------
Algorithm 2 interpolation is
------------------------------------------------
 input: vertexlist vl={{r, il_sub={j}}_i}
        indexlist il={j}
        index of vertex i
 output: interpolated vertex r'

 r' := 0 (r' := r_i)
 count := 0

 for each index j in vl(i).il_sub do
   case (il(j) mod 3) of
     0: di1 := -2, di2 := -1
     1: di1 := +1, di2 := +2
     2: di1 := -1, di2 := +1

     d1 := vl(j+di1).pos
     d2 := vl(j+di2).pos

     r' :=+ d1 + d2
     count :=+ 2 (count :=+ 3)

 return r' :=/ count
------------------------------------------------
```

### F. Roughness of a 3D curve

After Starting from a parametrically defined 3D-path given in cartesian coordinates with $N_{3D}$ points $p_i$, which are connected with $N-1$ direction-vectors $d_i = p_i+1 - p_i$, a value for the local curvature $\kappa_i$ can be calculated by determining the cosine of the angle $\alpha_i$ between adjacent normalized direction vectors $d_i/\|d_i\|$ and $d_{i+1}/\|d_{i+1}\|$ using the dot-product. The dot-product gives a range of $\{-1, 1\}$, whereby 1 means the direction-vectors are codirectional (parallel) and -1 means the direction-vectors are directional (anti-parallel). For the roughness $r_C$ of the entire 3D-path, the sum over all dot-products is calculated, whereby the value range is shifted to $\{0, 2\}$, which means higher value for increasing angle $\alpha_i$.

$$r_C = \frac{1}{N_{3D}-1} \sum_{i=1}^{N_{3D}-1} \left(1 - \frac{d_i \cdot d_{i+1}}{\|d_i\| \cdot \|d_{i+1}\|}\right)^n \quad (6)$$

The roughness $r_C$ is normalized by the number of cumulated angles $N_{3D}-1$. The exponent n defines the amount of penalty for high angles.

### G. 3D-path length

The length of a 3D curve $l_C$ is calculated by summing up the lengths of all direction vectors $\|d_i\|$ along the curve and can be calculated by

$$l_C = \sum_{i=1}^{N_{3D}-1} \|d_i\| \quad (7)$$

## II. RESULTS

For a first evaluation, a spiral phantom with identical size and parameters (2.5 windings with radius 30 mm and wire diameter of 6 mm) has been generated with different spatial resolution: sample 1 comes with an isotropic voxel size of 1.67 mm (60×60×60 voxel), sample 2 with an isotropic voxel size of 1.00 mm (100×100×100 voxel), sample 3 with an isotropic voxel size of 0.50 mm (200×200×200 voxel) and sample 4 with 1.67×0.50×1.00 mm³ (60×200×100 voxel). The extraction parameters and data processing was identical for all samples. In Fig. 7 the results are shown. With increasing spatial resolution (sample 1 to sample 3), the roughness of the sample (3D projection as well as 3D mesh) decreases. In the same way, the 3D path directly generated of the extracted centerline voxels shows a higher smoothness, which directly results in a more accurate CPR visualization. However, by applying the proposed correction based on the 3D mesh ($D_{sphere}$=20 mm), the roughness of the 3D path dramatically decreases and the CPR visualization is enhanced for all spatial resolutions.

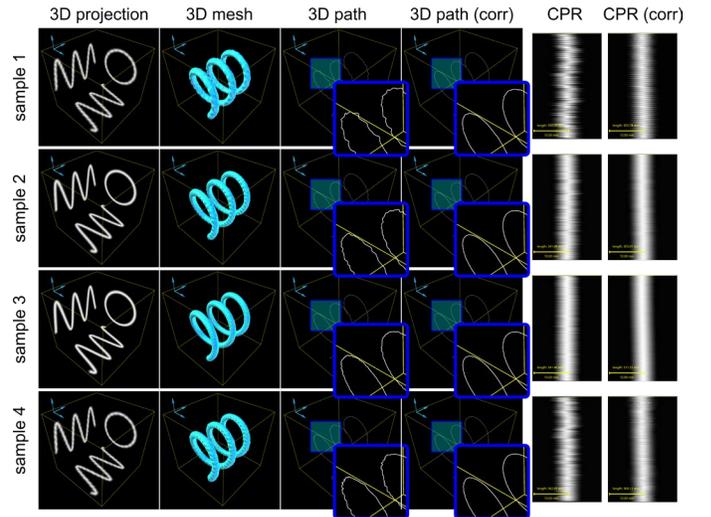

Fig. 7: Results of a simulated spiral sample with identical size but different spatial resolutions. With increasing spatial resolution (sample 1 to 3), the spiral shows more accuracy resulting in a smoother 3D path and CPR visualizations. Applying the proposed correction algorithm, the smoothness of the 3D path can dramatically be improved enhancing the CPR visualization.

The graph in Fig. 8 shows the roughness parameter $r_C$ depending on the sphere diameter $R_{sphere}$ ($n=1$) of a vessel sample (see Fig. 1). The inner diameter of the vessel structure is in the range of about $R_{vessel}$=1.5-2.5 mm ($D_{vessel}$=3-5 mm). When the applied diameter $D_{sphere}$=2·$R_{sphere}$ of the sphere is smaller than the vessel diameter, no correction can be applied and the $r_C$ parameter keeps constant. In a range between $D_{sphere}$=2.5-5.5 mm, the correction does not work for every segment point $s_i$ yielding large leaps and an increased $r_C$ parameter. However, the maximum of the peak can be found around $D_{sphere}$=4 mm, which corresponds quite well with the average diameter $D_{vessel}$ of the vessel structure.

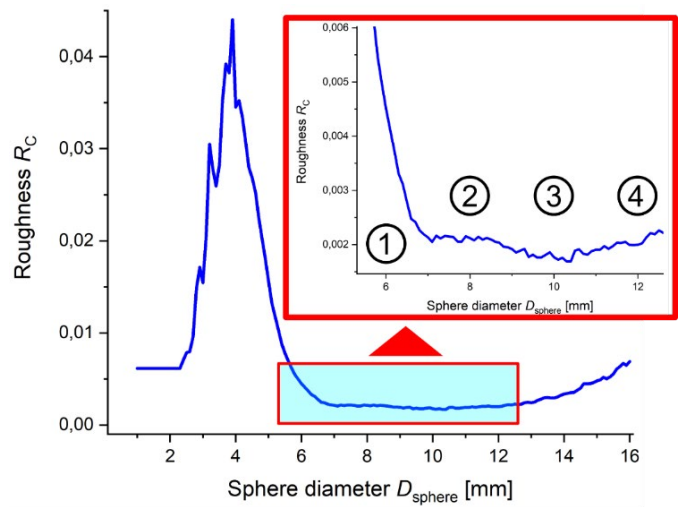

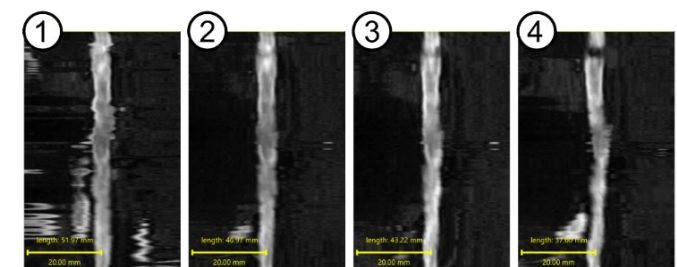

Fig. 8: **Top**: Graph shows the roughness $r_C$ in dependency of the sphere diameter $D_{sphere}$ for the vessel sample (Fig. 1). The maximum of the roughness corresponds with the average diameter of the vessel structure. **Bottom**: The CPR views for different sphere diameter {6 mm, 8 mm, 10 mm, 12 mm} are shown.

Between the sphere diameters $D_{sphere}$ of about 6 mm to 12 mm, the $r_C$ parameter keeps almost constant on a lower value. The corresponding CPR views in this range {6 mm to 12 mm} are shown in Fig. 8 (1)-(4). The optically best result can be found around 8 mm. With further increased sphere diameter $D_{sphere}$, the CPR image quality decreases due to interpolation effects.

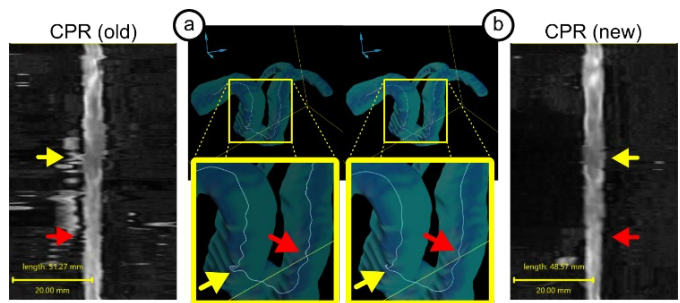

Fig. 9: **Top**: Results before **(a)** and after correction **(b)** show the optimized and smoothed 3D-path through the tortuous vessel structure (3D TOF MRI sequence). The new CPR view is optically enhanced and shows fewer artifacts in contrast to the not optimized CPR view.

Based on the results of Fig. 8, the parameter for the corrected CPR view was determined to $D_{sphere}$=7 mm ($N_{3D}$=500). In Fig. 9, the optimized 3D-path and enhanced CPR view (CPR new) is shown in contrast to the conventional processing. In contrast to the initial CPR view from Fig. 1 (CPR old), the roughness parameter could be decreased from $r_C$=0.00613 down to $r_C$=0.00215, which provides a smoother visualization with less artifacts.

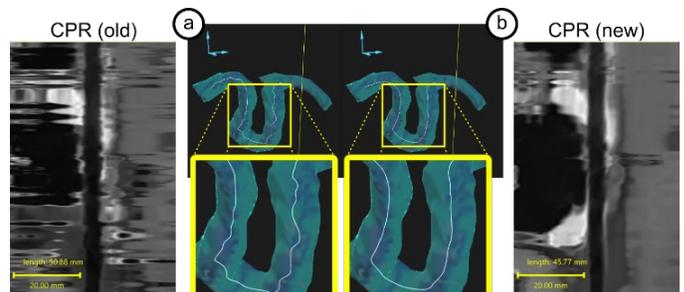

Fig. 10: Results before **(a)** and after correction **(b)** show the optimized and smoothed 3D-path through the tortuous vessel structure (CS-SPACE). The new CPR view is optically enhanced and shows fewer artifacts in contrast to the not optimized CPR view.

Fig. 10 shows the result with and without centerline optimization for the same vessel structure extracted from a CS-SPACE 3D MRI data set ($D_{sphere}$=8 mm, $N_{3D}$=500). The roughness parameter could be decreased from $r_C$=0.00432 down to $r_C$=0.00067.

III. DISCUSSION

The proposed algorithm utilizes MR images of the intracranial vasculature for improved luminal and mural assessment with fewer artifacts. In order to precisely measure mural thickness or luminal stenosis of an intracranial artery, e.g., when imaging vascular pathologies such as atherosclerosis and vasculitis, a true orthogonal view on the evaluated vessel segment can be generated from CPR produced with the presented algorithm.

Fig. 9 shows the improvement of the optimized centerline prediction. However, especially in the case of tortuous intracranial arteries, the curvature and small deviations in diameter can cause erroneous centerline predictions using a constant sphere radius over the entire vessel structure. Furthermore, in the case of strong curvature of the structure resulting in touching surfaces, a more sophisticated selection model is necessary. E.g., applying a dynamic sphere radius, based on the prior surface mesh information, or a cylindrical selection volume model based on a local centerline prediction can be used for a more accurate centerline prediction.

For centerline correction using a spherical approach as mentioned, an increasing sphere diameter causes a strong interpolation of the surface vertices, which smooths out the curvature of the tortuous vessel structure as shown in Fig. 8 (4).

In an extreme case, all available vertices are used for the interpolation process resulting in a single point (the length of the centerline is calculated to zero Eqn. (7)), which causes an increase of the roughness parameter. However, the calculation of the roughness parameter indicates the area for optimal sphere diameter (see Fig. 8 top). Furthermore, a diameter sweep not only provides the optimal parameters for correction, but also gives a good idea about the average diameter of the tubular structure.

A high contrast (positive or negative) is mandatory for a successful voxel extraction of the inner lumen. Often it is not possible to get a tubular structure due to SNR issues, especially in the intracranial regions of the human head since the signal-to-noise ratio is not sufficient for applying the extraction algorithms appropriately. This can cause inaccuracies in the surface mesh, such as holes or dents yielding erroneous centerline positions. To overcome this issue, multiple approaches are available for mesh restoration [18] providing finally a tubular structure.

In addition, the algorithm for generating the surface mesh from the binary 3D data set can be improved by more sophisticated meshing types like extended marching cube (EMC) [19] or marching tetrahedron [20].

Eqn. (1) and also algorithm 2 come along with two ways of interpolation, whereby one way is taking the initial position of the vertex $r_i$ into account. This can be helpful to stabilize the centerline prediction especially in cases of small vessels or data with low SNR.

## IV. Conclusion

The proposed approach provides an optimized method for the prediction of a centerline enhancing the visualization of tortuous vessel structure, such as intracranial arteries. Using a fast and robust way for centerline prediction in low-resolution MRI data set based on additional information about the surface mesh of the vessel provide an enhancement in image quality for curved planar reformation (CPR) views with less artifacts due to erroneous centerline. Furthermore, the average diameter of the vessel structure can be determined by applying a diameter sweep of the correction sphere. This technique allows more precise measurements of mural thickness or luminal stenosis of a tortuous intracranial artery by providing a true orthogonal view on the evaluated vessel segment.